# Denoising VAE as an Explainable Feature Reduction and Diagnostic Pipeline for Autism Based on Resting state fMRI

Xinyuan Zheng[3], Orren Ravid[2], Robert A.J. Barry[1], Yoojean Kim[2], Qian Wang[2], Young-geun Kim[1,2,5], Xi Zhu[1,2*], Xiaofu He[1,2,4*]

[1]Department of Psychiatry, Columbia University, New York, NY, United States
[2]The New York State Psychiatric Institute, New York, NY, United States
[3]Department of Statistics, Columbia University, New York, NY, United States
[4]Data Science Institute, Columbia University, New York, NY, United States
[5]Department of Biostatistics, Columbia University, New York, NY, United States

*Address correspondence to
Xiaofu He, PhD Xi Zhu, PhD
xh2170@cumc.columbia.edu; xi.zhu@nyspi.columbia.edu
Columbia University Department of Psychiatry /The New York State Psychiatric Institute
1051 Riverside Drive, New York, NY 10032

**Abstract** Autism spectrum disorder (ASD) is a range of developmental conditions characterized by restricted interests and difficulties in communication. The complexity of ASD has resulted in a deficiency of objective diagnostic biomarkers. Deep learning methods have gained recognition for addressing these challenges in neuroimaging analysis, but finding and interpreting such diagnostic biomarkers remain challenging computationally. Here, we propose an explainable feature reduction pipeline for resting-state fMRI (rs-fMRI). We used Ncuts parcellations (Craddock atlas) and Power atlas to extract functional connectivity data from rs-fMRI. By developing a denoising variational autoencoder (DVAE), our proposed method compresses the connectivity features into 5 latent Gaussian distributions, providing a low-dimensional representation of the data to promote computational efficiency and interpretability. To test the method, we employed the extracted latent representations to classify ASD using traditional classifiers such as support vector machine (SVM) on a large multi-site dataset. The 95% confidence interval for the prediction accuracy of the SVM is [0.63, 0.76] after site harmonization using the extracted latent distributions. Without using DVAE for dimensionality reduction, the prediction accuracy is 0.70, which falls within the interval. The DVAE encoded the diagnostic information from rs-fMRI data to 5 latent Gaussian distributions without sacrificing prediction performance. The runtime for training the DVAE and obtaining classification results from its extracted latent features was 7 times shorter compared to training classifiers directly on the raw data. Our findings suggest that the Power atlas provides more effective brain connectivity information for diagnosing ASD than Craddock atlas. Additionally, we visualized the latent representations to gain insights into the brain networks contributing to the differences between ASD and neurotypical brains.

## I. Introduction

Autism Spectrum Disorder (ASD) encompasses a spectrum of neurodevelopmental conditions presenting in early childhood. It is characterized by challenges in social interaction and communication, restricted interests, and repetitive behaviors. Clinical manifestations of ASD vary and include impairments in joint attention, eye contact, and sensory over-responsivity, among other characteristics [1]. Despite an estimated prevalence of approximately 1 in 32 children [2], ASD's heterogeneity renders diagnosis through direct pathological or radiological means challenging [3]. The importance of early identification to facilitate timely intervention is widely recognized, with the average age of diagnosis being around four and a half years [4], despite signs emerging as early as six months [5]. Diagnostic approaches vary and include observation- and interview-based methods such as the Childhood

Autism Rating Scale (CARS) as well as the Autism Diagnostic Interview-Revised (ADI-R) [6]. Manual methods, while comprehensive, require extensive time commitments, contain subjective elements, and are constrained by the necessity of expert interpretation [7].

Advancements in neuroimaging, particularly resting-state functional Magnetic Resonance Imaging (rs-fMRI), have provided novel insights into the neural underpinnings of ASD. These techniques offer objective biomarkers for early diagnosis using functional connectivity within the brain [8], [9]. Through the application of fuzzy spectral clustering [10], entropy analysis [11], and computational methods to evaluate spatial-temporal connectivity patterns, rs-fMRI has been used to distinguish ASD patients from typically developing individuals [12], [13]. Moreover, the advent of machine learning (ML) and deep learning (DL) technologies has enabled the analysis of complex rs-fMRI data, identifying patterns within functional connectivity that elude human observation. Such fMRI-based DL automated systems represent a promising adjunct for early ASD screening [13], although they are not yet sufficient for clinical use [7] and computationally expensive.

ML and DL models face challenges, particularly with small sample size and the high dimensionality of rs-fMRI feature vectors, which may contain tens of thousands of dimensions, thus complicating model training and interpretation. It has been observed that despite having large, multisite studies with over 2000 participants, models for predicting ASD tend to be brittle when facing dataset shift, which adversely affects the performance of DL models when applying to a difference dataset with a different data distribution [14]. Feature selection techniques such as Variational Autoencoders (VAE) [15], [16], support vector machine-recursive feature elimination (SVM-RFE) [17], and graph-based feature selection [18] have been utilized to enhance model accuracy in differentiating ASD from healthy controls. For ASD classification, supervised learning methods, such as SVMs, decision trees, and Gaussian naive Bayes, have been widely employed [19], [20]. Although convolutional neural networks are commonly used [21], they are limited by the non-Euclidean structure of functional connectivity matrices [22]. Innovations such as Deep Belief Networks (DBN) [23], Capsule Networks (CapsNet) [24], and ASD-Diagnet [15] have demonstrated prediction accuracies of 76%, 71%, and 70%, respectively, on the ABIDE (Autism Brain Imaging Data Exchange) dataset. Mellema et al. demonstrated accuracies exceeding 80% AUC employing a Dense Feed-Forward network on the IMPAC dataset [25], which is a superset of ABIDE formed by combining the ABIDE I, II datasets with an unpublished dataset from Robert Debré Hospital (RDB) in Paris, France.

Despite the developments in neuroimaging and ML and DL techniques, the development of models which are more interpretable, and less time consuming is paramount. As such, effective dimensionality reduction is a key challenge. A line of work has focused on studying the representations of rs-fMRI extracted by deep neural networks [21]. For instance, in the work by Liao and Lu [26], denoising autoencoders were trained with the NMI statistic matrix and representations were used to classify ASDs. Similarly, [10] and [27] applied stacked autoencoders, and [28] proposed multiple sparse autoencoders to extract low-dimensional features from rs-fMRI. These aforementioned works did not incorporate variational inference as used in variational autoencoders (VAEs), so the components in representations are typically not statistically independent, and the data log-likelihood may not be maximized. To overcome these limitations, [29] trained convolutional VAEs with rs-fMRI converted to polar coordinate space. Representations derived from VAEs were more effectively clustered by subjects. The VAE approach have shown potential in identifying conditions such as Alzheimer's disease [30], [31], Attention Deficit Hyperactivity Disorder (ADHD) [32], Post-Traumatic Stress Disorder (PTSD) [33], schizophrenia [34], [35] and ASD, albeit on a simpler dataset (ABIDE) with less sites and smaller age range [15]. Additionally, the challenges of model generalizability and interpretability have yet to be addressed.

Herein, we put forth a pipeline combining a Denoising Variational Autoencoder (DVAE) with ML classifiers, which demonstrates substantial dimensionality and computation reduction. We also implemented data harmonization and 5-fold cross-validation combined with adjusted threshold optimization. Employing this dimensionality reduction approach enhances computational efficiency while also mitigating the risk of overfitting. Moreover, low dimensional representations hold the potential to improve interpretability, and thus contribute to our understanding of ASD. To summarize this project, our work (1) built a DVAE to provide a low dimensional representation of the resting-state fMRI data, reducing the data size to 1/3500 and cutting the training time to 1/8, (2) evaluated the model's diagnostic performance with and without the dimensionality reduction model, (3) compared the performance of two brain atlases in diagnosing ASD based on functional connectivity information derived from each atlas, (4) tested model generalizability on a large multi-site dataset using leave-one-site-out cross-validation (LOSOCV), (5) tested the impact of demographic characteristics on the classification, and (6) provided a visualization for the encoded latent representations to enable network interpretability. Our code is publicly available online at https://github.com/xinyuan-zheng/Autism_DVAE.

## II. Methods

### *Participants*

To evaluate our proposed pipeline, we used a public dataset from Paris-Saclay Centre for Data Science that was initially published for competition in the Imaging-Psychiatry Challenge (IMPAC). The IMPAC dataset is generally considered to be a complex dataset due to the greater number of participants and a larger number of data collection sites (35 sites in IMPAC compared with 17 and 19 in ABIDE I and II, respectively). The dataset contained 1,150 subjects, comprising 601 healthy controls and 549 ASD patients. The dataset provider supplied image quality information by manually reviewing all time-series imaging data. 57 controls and 64 patients were marked as poor imaging quality. Following this review, our proposed pipeline was used to train, validate and test a total of 1,029 subjects. 80% of the data across all sites was randomly selected for training and validation, and the remaining 20% of the data was held out as the test set. The total number of subjects included in the training and validation set is 824, including 439 healthy controls and 385 ASD subjects. A total of 205 subjects were withheld as our test set, including 100 healthy controls and 105 ASD subjects. The age of the subjects ranged from 5 to 62 years. Table I summarizes demographics of the public dataset.

### *Processing: Functional connectivity of ROIs*

Researchers have found that in diagnosing mental disorders, functional brain parcellations yield better results than histological or anatomical features parcellations [36]. The functional brain parcellations are extracted from rs-fMRI, representing homogeneous parcels that perform actively in task-related brain activation [36]. By analyzing the correlation of brain function across these parcellations, it is possible to construct specific functional connectomes at both individual and group levels, enabling the differentiation of patients with ASD from healthy controls. The standard approach for processing rs-fMRI data involves computing a correlation matrix of pairwise correlations between brain regions. We extracted functional connectivity matrices from the preprocessed rs-fMRI data using the Power atlas [37], comprising 264 regions of interest (ROIs), and Ncuts parcellations (Craddock atlas) [36], comprising 249 ROIs. Both parcellations are functional brain atlases, providing connectivity information of the brain network. Each functional connectivity matrix extracted from the rs-fMRI imaging data is a symmetric matrix, where each element represents the Pearson correlation between two ROIs. These elements from the correlation matrices are then used as input features for further analysis. We vectorized the correlation matrix into a 1D correlation vector by the expansion of lower triangle values of the matrix, as shown in Figure 1A. For a general connectivity matrix with rows or columns, the length of the vectorized correlation vector will be $n \times (n+1)/2$. Accordingly, in our case, the 264 × 264 correlation matrix derived from the Power atlas results in a 34,980-dimensional vector, while the 249 × 249 correlation matrix from the Ncuts parcellations yields a 31,125-dimensional input vector. Additionally, to ensure the correlations are not driven by site-specific effects or confounding factors including age or gender, we performed a data harmonization technique known as the ComBat algorithm using neuroHarmonize [38].

### *Dimension Reduction: Denoising Variational Autoencoder (DVAE)*

Neuroimaging data in high-dimensional space often have latent lower dimensional representations, which may help elucidate the underlying structures. Representations are more effective and interpretable when each component serves distinct roles from all the others. VAEs have gained recognition for their ability to learn statistically independent latent factors, generate high-dimensional imaging data and perform effectively in various tasks in medical imaging applications.

VAEs [39] consist of two neural networks, an encoder and a decoder. The encoder maps observations into lower dimensional representations, and the decoder reconstructs these representations into the original observation space. The loss function used for VAEs, known as the negative evidence lower bound (ELBO), enforces the inverse relation between encoder and decoder and matches the distribution of features from the encoder and a user-specified prior distribution, e.g., a multivariate standard Gaussian distribution. By doing so, the latent representations can retain as much of the information as possible from the initial observations, while ensuring each component is statistically independent. To provide a detailed formulation, we denote encoder and decoder by $q_{\phi}(z|x)$ and $p_{\theta}(x|z)$, respectively, where $\theta$ and $\phi$ are network parameters. The VAEs are trained by minimizing negative ELBO: $loss(x,z) = -E_{z\sim q_{\phi}(z|x)}\ log\ \ p_{\theta}(x|z) + D_{KL}(q_{\phi}(z|x)||p(z))$, where $D_{KL}(p_1||p_2) = E_{z\sim p_1} log\left(\frac{p_1(z)}{p_2(z)}\right)$ denotes the Kullback-Leibler (KL) divergence and $p(z)$ denotes the prior distribution. Here, the first term is the reconstruction error, which enforces the inverse relationship between the encoder-decoder pair. The second term is the KL-regularization term, which ensures each component in the latent representation follows a $q_{\phi}(z|x)$ distribution and is independent of the others by matching $q_{\phi}(z|x)$ and $p(z)$.

In our study, we used a denoising VAE (DVAE) as a dimension reduction and feature selection method, which is a variation of the VAE with noise injected at the input [40]. Specifically, we applied the DVAE model to the functional connectivity matrices extracted from the rs-fMRI imaging data and injected Gaussian noise with a variance of 0.1 to the inputs in order to achieve robustness. The choice of 0.1 as the noise injection level was based on prior experiments [41]. The structure of the DVAE used is illustrated in Figure 1B. The layers in the network are fully connected. To avoid overfitting and expedite the training process, we constrained the number of the latent distributions to the size of 5, i.e., the model is forced to represent the connectivity matrices using 5 Gaussian distributions in the latent space. The size of the latent variables was tuned to be computationally economical without significantly compromising prediction accuracy. The model was implemented using PyTorch.

*Classification: machine learning classifiers*

For the classification task of detecting ASD, the latent representations extracted from the trained DVAE were used as features and input into the classical ML models, which are described in detail below. The overall processing and classification procedure is presented in Figure 1. Model building and grid searching of the best parameters for all classifiers was implemented using Scikit-learn [42]. The thresholds of the classifiers were tuned by maximizing the geometric mean of the sensitivity and specificity.

1. Random Forest

Random forest is an ensemble learning method that fits a number of decision trees and outputs the majority vote of the trees for a classification task. During the training phase, multiple subsets of the original dataset are randomly sampled, and each decision tree is constructed independently. The final prediction is obtained by aggregating the independent outputs of individual trees. Random forest exhibits several advantages as an ensemble meta-estimator, especially robustness to overfitting and the ability to handle high-dimensional data. The number of trees grown and the maximum depth of the decision trees were selected through grid search and cross-validation on the training data from the value sets [10, 50, 100, 500, 1000] and [1, 3, 5, 10, 20], respectively.

2. Support Vector Machine (SVM)

SVM is an algorithm that finds the optimal hyperplane, using the maximum margin principle, to effectively separate data belonging to different categories in a higher-dimensional space. In the context of SVM, the support vector denotes the data points that lie closest to the decision boundary, and the algorithm focuses on maximizing the margin between these points. SVM algorithm is known for its efficiency and robustness for classification tasks that involve high-dimensional data. The choice of the kernel function determines how the input features undergo transformation to establish the decision boundary. The regularization parameter governs the trade-off between achieving a large margin and minimizing misclassifications. For a nonlinear kernel, the kernel coefficient gamma controls the radius of influence of the training samples. In our case, the choice of kernel between linear and radial basis function (RBF), was determined through grid search and cross-validation. The regularization parameter C and kernel coefficient $\gamma$ classifier were also selected by grid-search within the value sets C = [0.01, 0.1, 1, 10, 100] and $\gamma$ = [1, 0.1, 0.01, 0.001, 0.0001], respectively.

*Permutation Test*

A Permutation test is a non-parametric statistical method that can be used to assess the performance of a model and provide insights into the reliability of the achieved outcomes. The procedure begins by computing the performance metric of the model on the original dataset. Subsequently, the labels of the dataset are randomly permuted, and the performance metric is recalculated for each permutation. This step is iterated to create an empirical distribution of the objective metric. By comparing the observed actual metric to the distribution generated by randomizing labels, the model's performance can be statistically tested for significance, providing a robust approach for evaluating the effectiveness of a model beyond chance levels. In our study, we applied the permutation test to our ML classifiers. We shuffled the labels of the extracted latent representations and retrained the classifiers 1,000 times. The p-values of the classifier performances were computed based on the permutation iterations.

# III. Results

1. ML Classification performance based on DVAE latent features

The classification results are shown in Table II and III, where the 95% confidence intervals of the classification accuracy on the holdout dataset are reported. We found that using the Power atlas with 264 ROIs and SVM classifier, applying a ComBat algorithm to adjust for site, age, and gender covariates, and leveraging the DVAE to reduce feature dimensionality, gave the best performance on the test dataset. Figure 2 shows the loss curve of the DVAE, from which we conclude the convergence of the model. SVM (67% mean accuracy) was observed to perform better than random forests (62% mean accuracy) when classifying ASD patients from neurotypical controls. Figure 3 shows the AUC of the two classifiers. We also performed a permutation test of 1,000 iterations for the SVM and random forest classifiers, as shown in Figure 4. The accuracy score of 0.67 of a SVM yields a p-value of 0.0001, while the accuracy score of 0.62 of a random forest classifier yields a p-value of 0.003.

2. Effects of dimensionality reduction

The DVAE encoded the 34,980 dimensional input vector of the rs-fMRI connectivity matrix into 5 Gaussian distributions, corresponding to a 10-dimensional latent representation (each Gaussian distribution is parameterized by a mean and a variance). Without feature reduction, the classification performance is presented in Table IV, including classification AUC, accuracy, sensitivity and specificity scores. Using the rs-fMRI connectivity data, we reported an accuracy of 70% for SVM and 61% for random forest. The AUC and accuracies achieved using all functional connectivity data features fall within the 95% confidence interval of those achieved by the DVAE latent representations, indicating the effectiveness of DVAE in finding a compressed representation.

We find that the DVAE enabled ML classifiers to achieve comparable results with far less data input, and that SVM generally outperformed random forest. Notably, this reduction in both feature complexity and computation burden allows us to use bootstrapping methods to evaluate the ML classifiers. After dimensionality reduction using the Power atlas and the DVAE model, SVM achieved a 95% confidence interval of [0.63, 0.76] for prediction accuracy and RF gave a test accuracy confidence interval of [0.53, 0.66]. In contrast, classification using raw features was computationally expensive, making it infeasible to assess confidence intervals.

Moreover, we observed a significant reduction in computational time when adopting the DVAE approach. Training a single DVAE model to extract latent representations had a runtime of 35 minutes, while running classifiers on the compressed latent representations with grid search for optimal parameters took less than 2 minutes. In comparison, training ML classifiers directly on the connectivity matrices with a grid search for optimal parameters required a runtime of 5 hours in our experiments.

3. Comparison between the Ncuts parcellations and Power atlas

As shown in Supplementary Table I, with Combat, SVM gave an undesirable 95% confidence interval of [0.48, 0.64] using DVAE latent representations extracted from Ncuts parcellations during cross validation, and random forest gave a 95% confidence interval of [0.62, 0.75] using DVAE latent representations extracted from Ncuts parcellations. Without Combat, the performance of both classifiers was slightly worse. SVM gave a 95% confidence interval of [0.48, 0.62], and random forest gave a 95% confidence interval of [0.59, 0.73]. Our results show that the Power atlas offers more efficient brain connectivity information for ASD detection compared to Ncuts parcellations.

### 4. Leave-one-site-out cross-validation (LOSOCV)

As the dataset consists of 35 different data acquisition sites, the number of subjects at most sites is insufficient for a complete leave-one-site-out cross-validation (LOSOCV). As a result, we validated the performance of the ML classifiers on the 4 sites that collected data from more than 20 neurotypical controls and 20 ASD patients. Table V presents the performance of the models tested on these qualified sites. For SVM, the average AUC is 0.63, and the average accuracy is 0.63. For random forest, the average AUC is 0.60 and the average accuracy is 0.63. The model performed similarly across sites with small sample sizes.

### 5. Effects of age and gender

One distinguishing characteristic of the IMPAC dataset is its wide age range, with subjects ranging from 5 to 64 years old, along with an unbalanced sex distribution. We applied a ComBat algorithm to adjust for age and gender effects. As indicated by previous studies [43], [44], there are sex differences in the functional organization of the brains of individuals with ASD. Hence, to test the generalizability of our method, we evaluated the effects of these covariates by concatenating and inputting them along with the latent features as an additional piece of information to ML classifiers. The inclusion of age and gender as additional inputs to the model resulted in performance similar to classifiers using only DVAE-compressed features. The additional information did not impact the classification performance. The quantitative results reported are in Table VI, VII and VIII. The AUC plots for models with additional inputs are presented in Figure 5.

### 6. Visualization and interpretation of the latent representations

To interpret the latent representations of the DVAE, we utilized latent contribution scores (LCS) to quantify the contribution of each latent representation $z$ on the reconstructed $\hat{x}$[41]. Defined as $E_{q_{\phi}}(z|x)\ \partial\hat{x}/\partial z$, LCS is an input perturbation-based feature importance measure that estimates the marginal changes in the reconstructed results with respect to the latent representations. This approach enables the location and visualization of the functional connectivity exhibiting the highest LCS for each latent representation, as illustrated in Figure 6. The network connectivity and brain regions exhibiting the highest LCS are reported in Supplementary Figure 1 and 2. To visualize the latent representations, we reconstructed them using the decoder of the trained DVAE. Samples from two representative subjects are displayed in Supplementary Figure 3. Similarly, Supplementary Figure 4 presents the aggregated reconstructed latent representations for the Autism and healthy control groups, along with the differences between the groups for each latent representation.

## IV. Discussion

In this study, we developed a hierarchical feature reduction pipeline that leverages functional brain patterns. Neuroimaging studies often face the challenge of high data dimensionality relative to the sample size. To address this, our study employed a DVAE for dimensionality reduction, combined with classical ML classifiers, to derive insights into ASD using rs-fMRI data. Using a large, public, multi-site dataset, we validated the effectiveness and generalizability of our pipeline across multiple sites. The classifiers achieved a 95% confidence interval of [0.63, 0.76] for accuracy and [0.60, 0.74] for AUC on independent hold-out data after site harmonization using latent representations encoded from functional connectivity data. Our results also indicate that the Power atlas provides more efficient brain connectivity information for ASD diagnosis. To test the generalizability of our method, we evaluated the performance of the model using a LOSOCV approach and analyzed the impact of demographic characteristics by incorporating age and sex as covariates in the classifiers. The inclusion of demographic information led to marginal changes in classification performance, suggesting potential commonalities in the functional representations of ASD across different age groups and genders.

Overall, the classifiers achieved comparable results for the diagnosis of ASD using encoded latent representations with approximately 1/3500 the number of the original features from the rs-fMRI functional connectivity. The DVAE model effectively encoded the functional brain imaging in terms of classifying ASD patients and healthy controls. The proposed pipeline also mitigated the overfitting problem observed with traditional ML classifiers. Moreover, dimensionality reduction accelerated the training of the ML classifiers and testing for new subjects, enabling statistical evaluation of classification results in a computationally feasible manner. The model provided quick runtime and generalizability for multi-site data, making it more adaptable and practical to diverse clinical settings in practice.

Compared to subjective assessments using DSM-5, automatic methods based solely on neuroimaging have lower diagnostic accuracy but may provide valuable insights into the brain regions affected by the disease. Beyond diagnosis, encoding methods hold promise for advancing our understanding of the disease mechanism. In the context of detecting biomarkers for ASD, the DVAE model provides a novel perspective by encoding differences in brain connectivity between patients and healthy controls into a latent space. This approach enables the interpretation of extracted latent representations by computing the network distribution of the latent features and visualizing them on brain maps [41]. Such visualization facilitates the identification of brain regions and connections affected by the disorder, thereby allowing the functional underpinnings of ASD to be better understood by clinical experts and researchers. The potential of using fMRI as a method to diagnose ASD and predict disease progression also needs to be investigated further in additional studies.

## V. Conclusion

We employed a DVAE model for dimensionality reduction alongside machine learning classifiers to detect ASD using rs-fMRI data. Our approach developed a hierarchical feature reduction pipeline that leverages functional brain patterns. The DVAE model encoded the diagnostic information from rs-fMRI data, reducing over 30,000 features to 1/3500 of the original size. Additionally, we found Power atlas offers more effective brain functional connectivity information for diagnosing ASD compared to the Craddock atlas. The latent representations of the rs-fMRI data were visualized to provide insights into ASD. The DVAE model offers a novel perspective for the understanding of brain connectivity differences between patients and healthy controls.

**Statements & Declarations**

- Ethics approval and consent to participate

Not applicable.

- Data and code availability

Data used in this study is a public dataset. The code is open source to ensure reproducibility and can be accessed online at https://github.com/xinyuan-zheng/Autism_DVAE.

- Competing interests

The authors declare no competing interests.

- Authors' contributions

XYZ contributed to the investigation, implementation, and visualization of the model, data processing, statistical analysis, and drafting of the manuscript. OR contributed to the implementation and visualization. RAJB contributed to the model investigation and manuscript drafting. YK and QW contributed to data processing and model investigation. YGK contributed to the review and editing of the manuscript. XZ and XH contributed to the supervision of the study, as well as the editing of the manuscript. All authors participated in manuscript revisions and approved the final draft.

- Funding

This study was supported by the NIH K01MH122774 and the Brain and Behavior Research Foundation NARSAD Young Investigator (XZ).

Table I: Demographics of the dataset.

| Site | Subject | M | F | Age Mean ±SD [Min, Max] | Healthy Control | ASD |
|---|---|---|---|---|---|---|
| 0 | 39 | 39 | 0 | 39.18 ±14.41[18.00, 62.00] | 17 | 22 |
| 1 | 19 | 15 | 4 | 8.08 ±1.05 [6.33, 10.65] | 10 | 9 |
| 2 | 34 | 23 | 11 | 10.49 ±1.71 [8.06, 13.88] | 21 | 13 |
| 3 | 27 | 12 | 15 | 22.73 ±11.40 [7.56, 46.6] | 20 | 7 |
| 4 | 16 | 10 | 6 | 26.13 ±9.97 [17.00, 54.00] | 10 | 6 |
| 5 | 97 | 60 | 37 | 10.42 ±1.24 [8.07, 12.99] | 72 | 25 |
| 6 | 21 | 21 | 0 | 23.38 ±3.81[18.00, 33.00] | 0 | 21 |
| 7 | 48 | 43 | 5 | 10.11 ±5.98 [5.22, 34.76] | 16 | 32 |
| 8 | 12 | 9 | 3 | 6.47 ±1.04 [5.13, 8.84] | 0 | 12 |
| 9 | 54 | 33 | 21 | 11.50 ±2.03 [8.00, 15.00] | 28 | 26 |
| 10 | 31 | 25 | 6 | 13.27 ±3.01 [7.40, 18.00] | 12 | 19 |
| 11 | 20 | 20 | 0 | 15.78 ±2.83 [12.00, 20.00] | 12 | 8 |
| 12 | 18 | 13 | 5 | 15.01 ±1.67 [12.08, 17.42] | 8 | 10 |
| 13 | 18 | 14 | 4 | 23.09 ±7.45 [13.55, 38.86] | 10 | 8 |
| 14 | 19 | 13 | 6 | 29.16 ±10.92 [18.70, 56.20] | 11 | 8 |
| 15 | 13 | 9 | 4 | 26.15 ±6.15 [19.00, 40.00] | 6 | 7 |
| 16 | 30 | 23 | 7 | 10.45 ±1.28 [8.09, 12.76] | 16 | 14 |
| 17 | 18 | 18 | 0 | 21.89 ±2.66 [18.00, 29.00] | 10 | 8 |
| 18 | 16 | 12 | 4 | 14.41 ±1.40 [12.30, 16.90] | 10 | 6 |
| 19 | 33 | 27 | 6 | 28.76 ±11.28 [7.00, 52.00] | 21 | 12 |
| 20 | 101 | 77 | 24 | 15.48 ±6.97 [7.13, 39.10] | 58 | 43 |
| 21 | 11 | 11 | 0 | 10.59 ±1.38 [8.20, 12.65] | 8 | 3 |
| 22 | 21 | 18 | 3 | 17.19 ±3.59 [10.00, 24.00] | 9 | 12 |
| 23 | 32 | 26 | 6 | 19.58 ±7.18 [9.33, 35.20] | 14 | 18 |
| 25 | 12 | 12 | 0 | 35.50 ±4.75 [27.00, 42.00] | 5 | 7 |
| 26 | 22 | 16 | 6 | 14.66 ±1.98 [8.67, 17.15] | 14 | 8 |
| 27 | 16 | 14 | 2 | 10.14 ±1.58 [7.75, 12.43] | 9 | 7 |
| 28 | 30 | 30 | 0 | 17.65 ±3.73 [12.25, 25.91] | 13 | 17 |
| 29 | 34 | 28 | 6 | 13.03 ±2.22 [8.49, 17.78] | 17 | 17 |
| 30 | 11 | 10 | 1 | 12.28 ±1.14 [10.04, 13.63] | 5 | 6 |
| 31 | 50 | 33 | 17 | 13.65 ±2.88 [8.20, 18.90] | 33 | 17 |
| 32 | 26 | 24 | 2 | 15.94 ±3.71 [12.80, 28.80] | 14 | 12 |
| 33 | 56 | 56 | 0 | 22.84 ±7.78 [9.95, 50.22] | 23 | 33 |
| 34 | 24 | 19 | 5 | 12.92 ±3.11 [7.00, 17.83] | 12 | 12 |
| **All** | **1029** | **813** | **216** | **17.58 ±9.47 [5.13, 62.00]** | **544** | **485** |

Table II: Classification accuracy using DVAE latent representations.

| Accuracy 0.95 CI | With Combat | Without Combat |
|---|---|---|
| SVM | [0.63, 0.76] | [0.61, 0.75] |
| RF | [0.53, 0.66] | [0.48, 0.63] |

Table III: Classification AUC using DVAE latent representations

| AUC 0.95 CI | With Combat | Without Combat |
|---|---|---|
| SVM | [0.60, 0.74] | [0.61, 0.74] |
| RF | [0.52, 0.65] | [0.47, 0.68] |

Table IV: Classification performance using all connectivity features.

| | Sensitivity | Specificity | Accuracy | AUC |
|---|---|---|---|---|
| **SVM** | 0.70 | 0.67 | 0.70 | 0.72 |
| **RF** | 0.55 | 0.71 | 0.61 | 0.61 |

Table V: Random forest and SVM LOSOCV results using DAVE latent representations.

| | | Sensitivity | Specificity | Accuracy | AUC |
|---|---|---|---|---|---|
| **SVM** | Site 5 | 0.64 | 0.51 | 0.55 | 0.61 |
| | Site 9 | 0.58 | 0.64 | 0.61 | 0.6 |
| | Site 20 | 0.53 | 0.78 | 0.67 | 0.65 |
| | Site 33 | 0.79 | 0.57 | 0.7 | 0.66 |
| | **Avg. across sites** | **0.63** | **0.62** | **0.63** | **0.63** |
| **RF** | Site 5 | 0.64 | 0.5 | 0.54 | 0.54 |
| | Site 9 | 0.85 | 0.43 | 0.63 | 0.56 |
| | Site 20 | 0.53 | 0.83 | 0.7 | 0.69 |
| | Site 33 | 0.55 | 0.78 | 0.64 | 0.63 |
| | **Avg. across sites** | **0.64** | **0.63** | **0.63** | **0.6** |

Table VI: Test performance using DVAE latent representations and age and sex information.

| | Sensitivity | Specificity | Accuracy | AUC |
|---|---|---|---|---|
| **SVM** | 0.66 | 0.65 | 0.65 | 0.70 |
| **RF** | 0.63 | 0.65 | 0.63 | 0.65 |

Table VII: Test performance using DVAE latent representations and age information.

| | Sensitivity | Specificity | Accuracy | AUC |
|---|---|---|---|---|
| **SVM** | 0.69 | 0.62 | 0.65 | 0.67 |
| **RF** | 0.71 | 0.54 | 0.62 | 0.65 |

Table VIII: Test performance using DVAE latent representations and sex information.

| | Sensitivity | Specificity | Accuracy | AUC |
|---|---|---|---|---|
| **SVM** | 0.70 | 0.62 | 0.66 | 0.69 |
| **RF** | 0.70 | 0.56 | 0.63 | 0.66 |

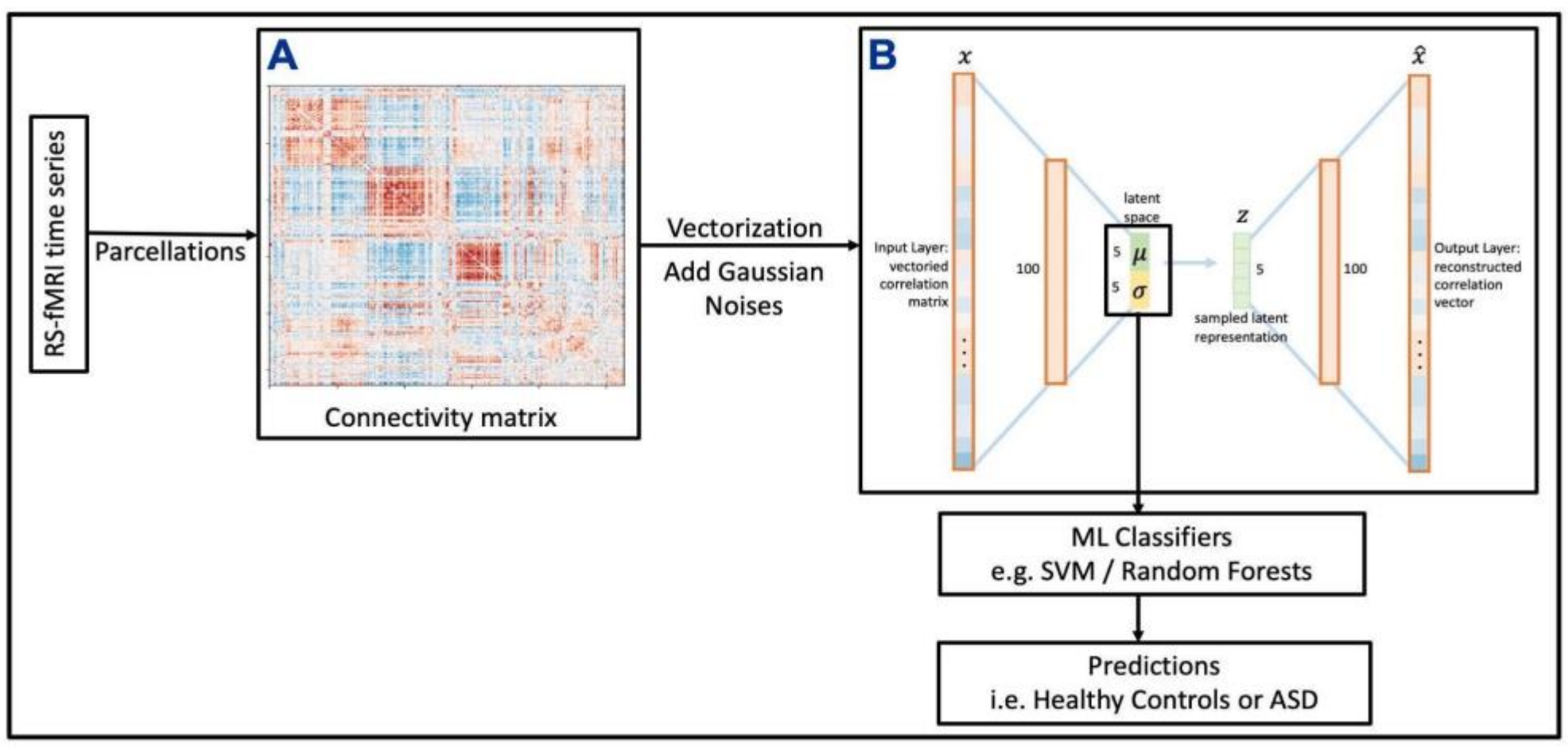

Figure 1: Proposed DVAE+ML classifiers Pipeline. The connectivity matrix was extracted from RS-fMRI data using brain parcellations for each subject. Gaussian noise was injected into the vectorized connectivity data and then input into the DVAE model. The model extracts latent features from the input, which are then used for predictions with traditional machine learning classifiers.

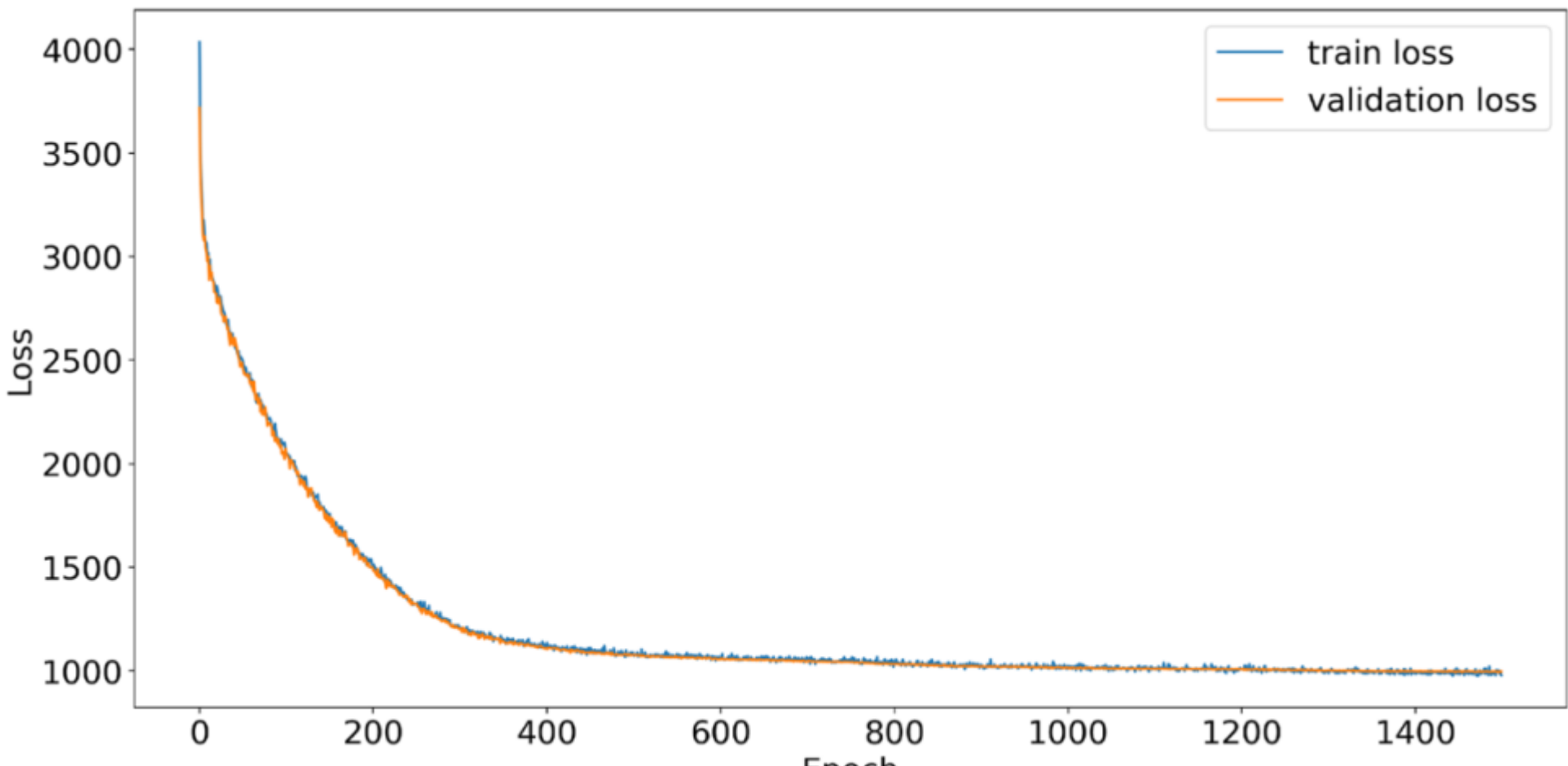

Figure 2: Loss versus epoch. The model converged as the loss function consistently decreased throughout the training process.

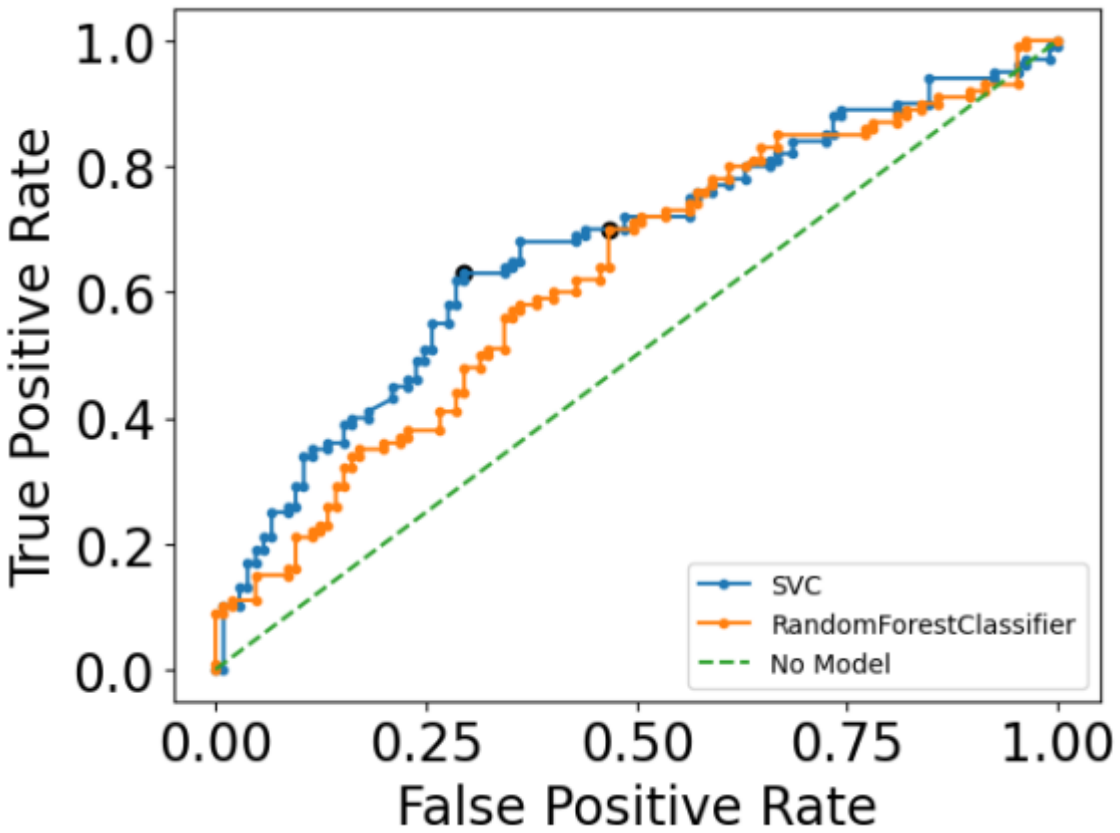

Figure 3: AUC-ROC plot of an experiment using DVAE extracted latent representations. Blue line shows the performance of one SVM classifier using the latent distributions as input. Orange line shows the performance of one random forest classifier using the latent distributions as input.

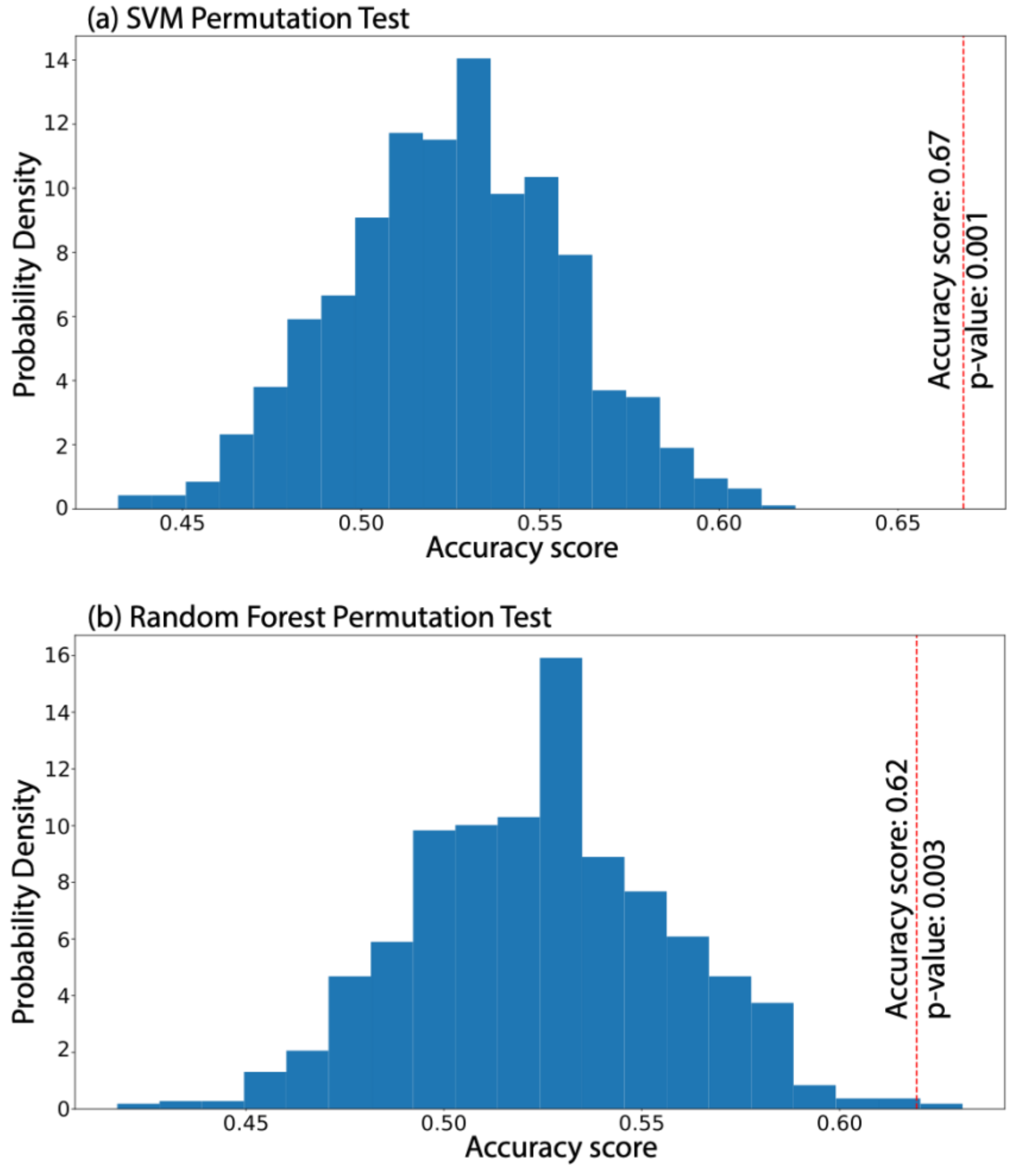

Figure 4: The results of permutation tests for SVM and random forest classifiers. Red dotted line shows the significance of the accuracy achieved by the proposed models.

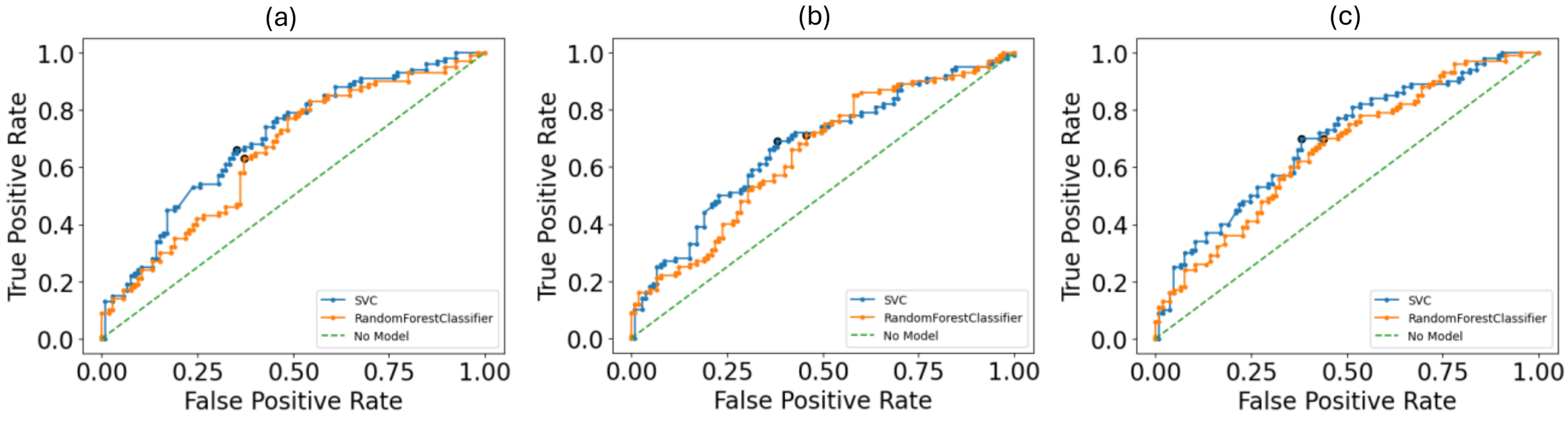

Figure 5: Prediction results using (a) age and gender, (b) age, (c) sex as additional information for the classifier.

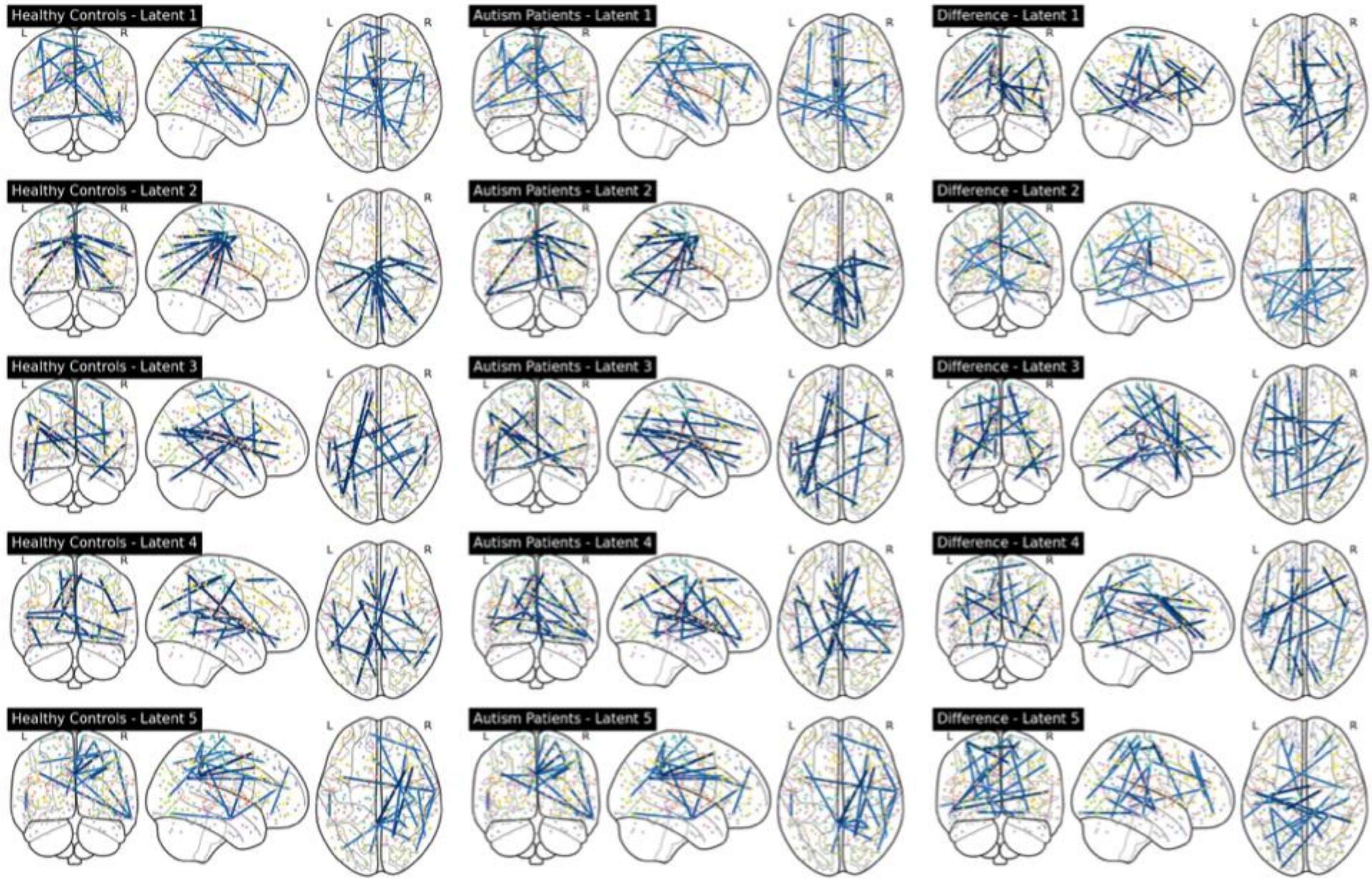

Figure 6: Top functional connectivity based on latent contribution scores.

## Supplementary material

Supplementary material related to this article can be found in the online version.

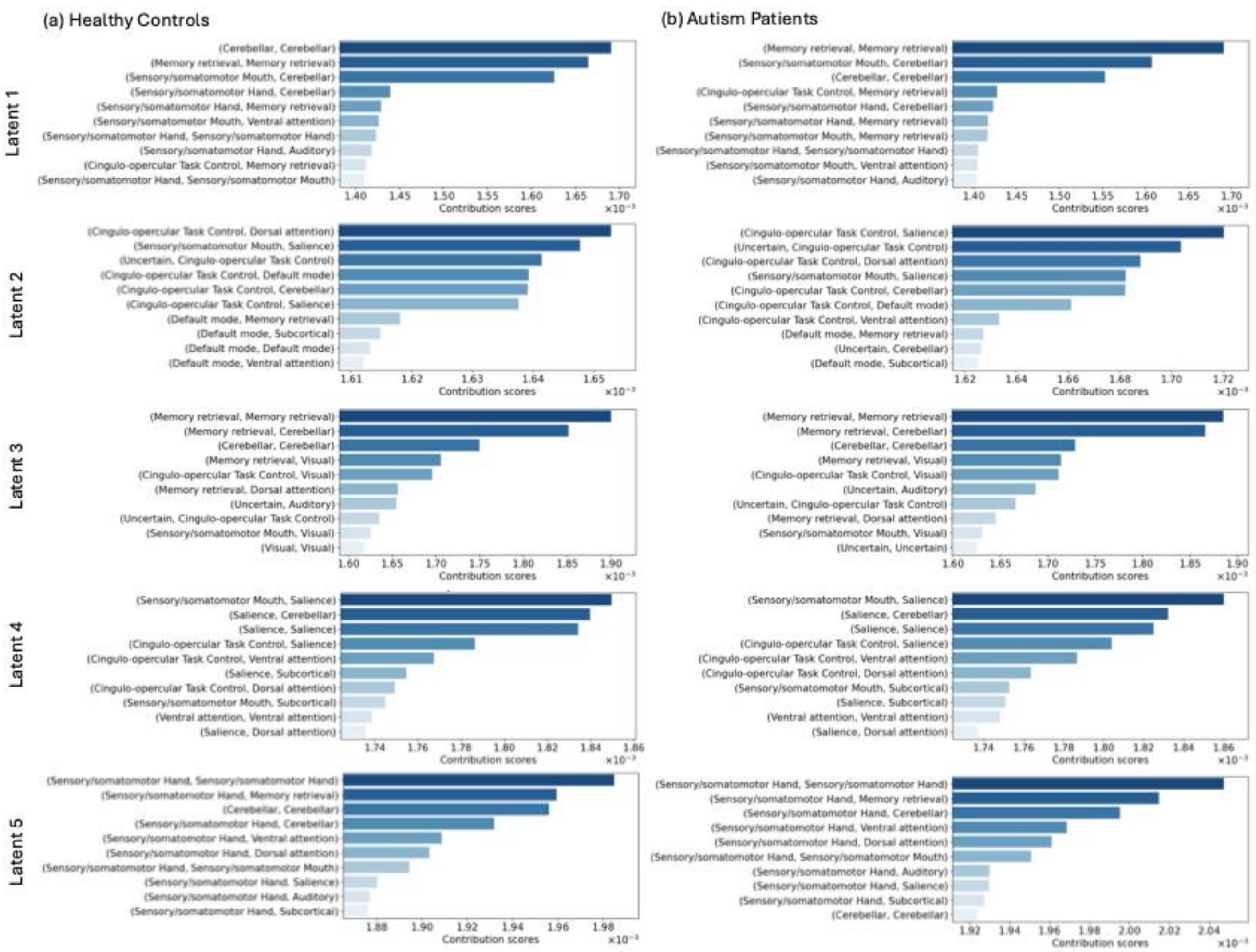

Supplementary Figure 1: The network connectivity exhibiting the highest LCS for each latent representation.

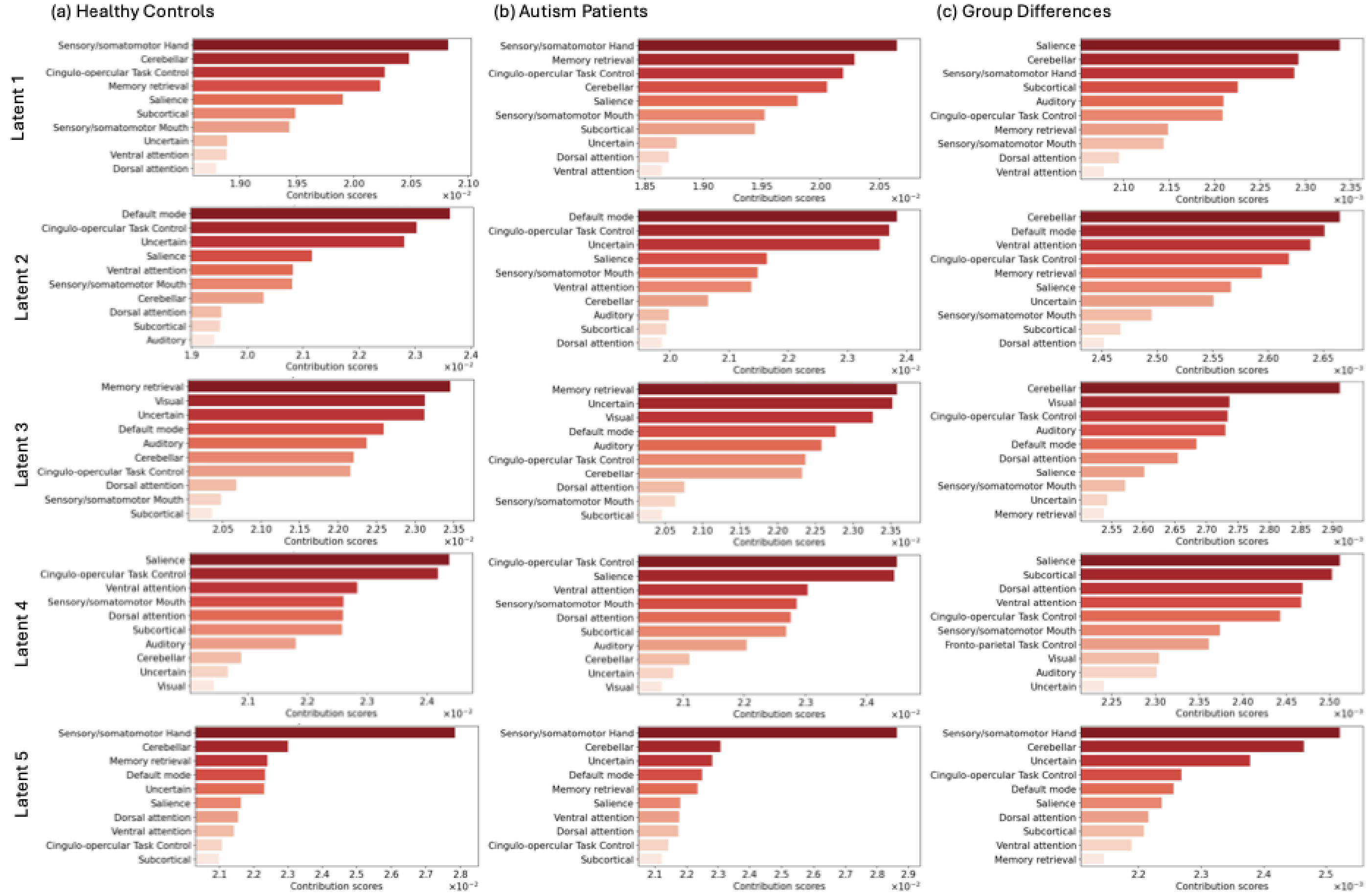

Supplementary Figure 2: The brain regions exhibiting the highest LCS for each latent representation.

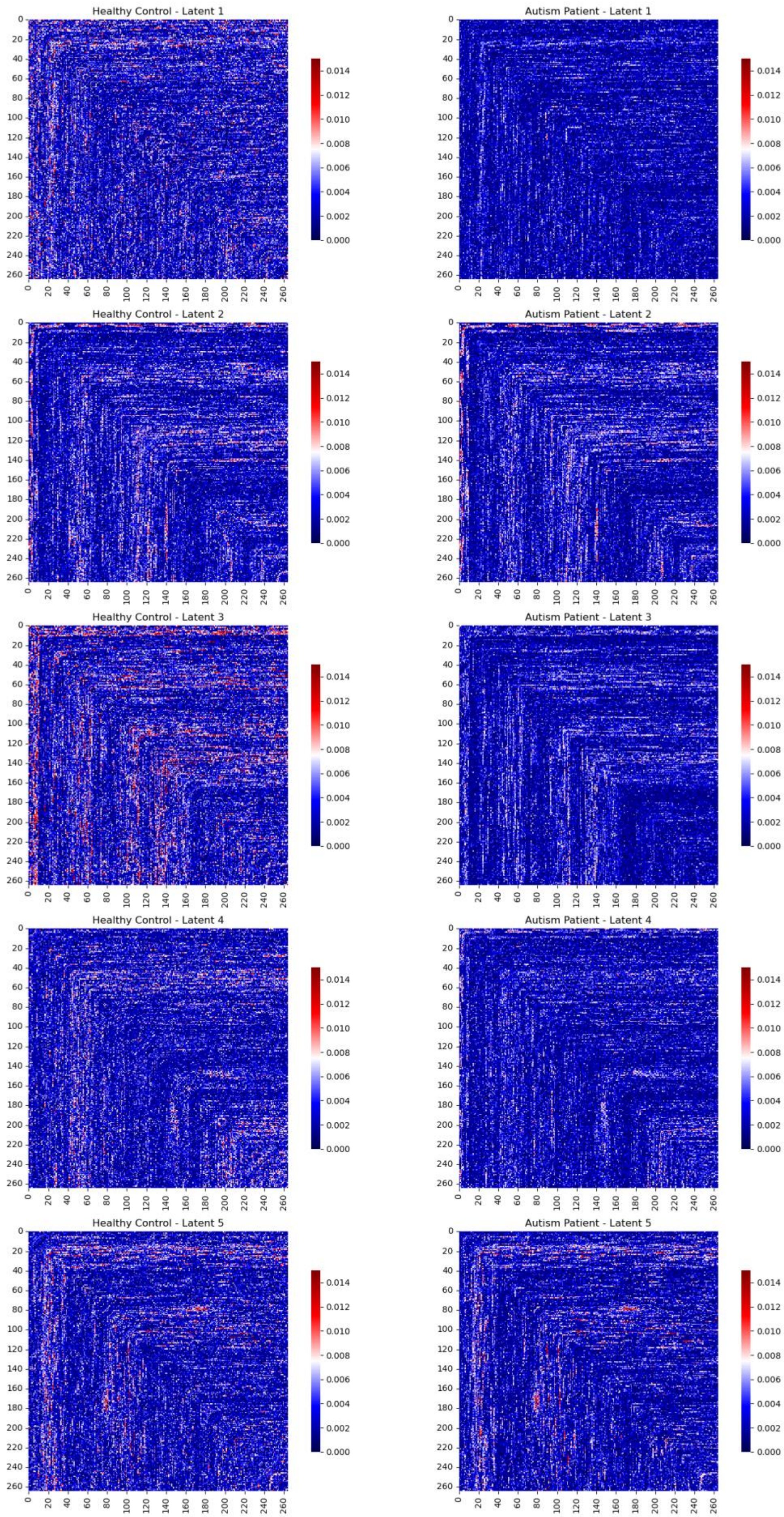

Supplementary Figure 3: Representative samples of decoded latent variables.

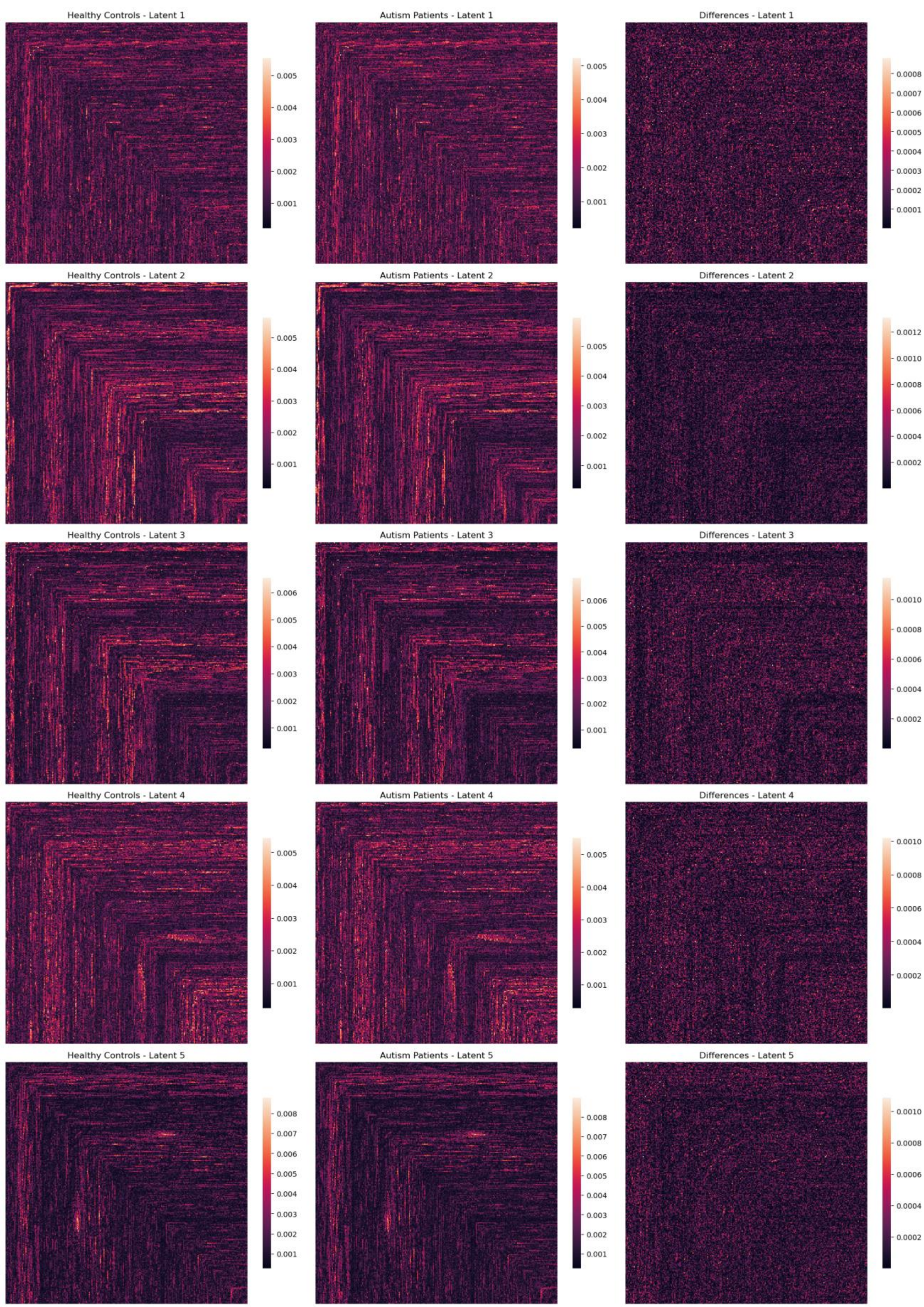

Supplementary Figure 4: Decoded latent representations of healthy controls, autism patients and the differences between the two groups.

Supplementary Table I: Test performance using DVAE latent representations extracted from Ncuts parcellations (Craddock atlas).

| **Accuracy 0.95 CI** | **With Combat** | **Without Combat** |
|---|---|---|
| SVM | [0.48, 0.64] | [0.48, 0.62] |
| RF | [0.62, 0.75] | [0.59, 0.73] |